\documentclass[aps, pra, superscriptaddress, twocolumn, amsfonts, amsmath, amssymb, floatfix]{revtex4-2}
\usepackage{graphicx}
\usepackage{float}
\usepackage{sidecap}
\usepackage{dcolumn}
\usepackage{bm}
\usepackage{epsfig}
\usepackage{braket}
\usepackage{color}
\usepackage{amsmath}
\usepackage[colorlinks=true, letterpaper=true, pdfstartview=FitV, linkcolor=blue, citecolor=blue, urlcolor=blue]{hyperref}

\begin{document}

\title{Adjusting exceptional points using saturable nonlinearities}

\author{Qingxin Gu}
\affiliation{Department of Physics and Institute for Quantum Science and Technology, Shanghai University, Shanghai 200444, China}

\author{Chunlei Qu}
\email{cqu5@stevens.edu}
\affiliation{Department of Physics, Stevens Institute of Technology, Hoboken, NJ 07030, USA}
\affiliation{Center for Quantum Science and Engineering, Stevens Institute of Technology, Hoboken, NJ 07030, USA}

\author{Yongping Zhang}
\email{yongping11@t.shu.edu.cn}
\affiliation{Department of Physics and Institute for Quantum Science and Technology, Shanghai University, Shanghai 200444, China}

\begin{abstract}
We study the impact of saturable nonlinearity on the presence and location of exceptional points in a non-Hermitian dimer system. The inclusion of the saturable nonlinearity leads to the emergence of multiple eigenvalues, exceeding the typical two found in the linear counterpart. To identify the exceptional points, we calculate the nonlinear eigenvalues both from a polynomial equation for the defined population imbalance and through a fully numerical method. Our results reveal that exceptional points can be precisely located by adjusting the non-equal saturable nonlinearities in the detuning space. 
\end{abstract}

\maketitle

\section{Introduction}
Recent advancements in the study of non-Hermitian systems have demonstrated a rich physics distinct from that of their Hermitian counterparts. Unlike Hermitian systems, where the degenerate eigenvectors are orthogonal with each other, non-Hermitian systems feature the emergence of exceptional points (EPs). These points are intriguing singularities where eigenvalues and their corresponding eigenvectors coalesce simultaneously~\cite{Bender1998,Berry2004,Miri2019}. Besides their theoretical significance, more and more functional applications related to EPs have been discovered, including chirality phenomena~\cite{Feng2022}, unidirectional reflection~\cite{Ramezani2010}, and EP-enhanced sensing in optical systems or classical electric circuits~\cite{Hodaei2017,Rouhi2022}, among others.

If a non-Hermitian Hamiltonian is pseudo-Hermitian \cite{Mostafazadeh2002}, its eigenvalues may be real-valued under certain conditions~\cite{Bender1998,Bender1999,jinliang2023P}. The widely investigated pseudo-Hermitian Hamiltonians are often characterized by parity-time (PT) symmetry with balanced gain and loss~\cite{Bender2007}. These systems may undergo a spontaneous phase transition, where eigenvalues evolve from real to complex beyond a critical point~\cite{Makris2008}. The critical point, known as an EP, separates the PT-symmetric phase from the PT-broken phase. Such phase transitions at EPs have stimulated further research interest in EPs-related physics. Recently, EPs have been experimentally observed in various physical systems~\cite{Miri2019, Klaiman2008,Ruter2010,Bender2013,lee2009}.

Given that most of these platforms are nonlinear systems, the effect of nonlinearity on EPs and the related phase transitions has soon gained attention~\cite{Konotop2016}. It has been found that nonlinearity can shift the location of EPs in the parameter space of a PT-symmetric periodic system~\cite{Lumer2013,Zhang2021}. A nonlinear dimer, being the simplest system with both PT symmetry and nonlinearity, has seen extensive investigation into the bifurcation of EPs from nonlinear modes~\cite{Graefe2006,Graefe2008,Graefe2010,Graefe2012,Suwunnarat2020,Li2011,Dast2013_1,Dast2013_2,Wu2022}. Nonlinearity has also been shown to play a crucial role in functional applications, such as modulation of EPs~\cite{Chen2022}, gyroscopes~\cite{Zhang2022}, \textcolor{red}{non-hermitian electrical lines~\cite{non-hermitian electrical lines}}, and increasing the order of EPs~\cite{Bai2023Xiao}, as well as inducing a non-reciprocal nonlinear Landau-Zener tunneling in a dynamically accelerated nonlinear dimer~\cite{Wang2022}.

EPs are singular points in parameter space, around which eigenvalues change abruptly, leading to profound applications such as mode selection and sensitivity enhancement  ~\cite{Hodaei2017,Xiao2019,jinliang2020High-orderEP,jinliang2023singularity}. However, the precise control of parameters and the sharp changes in physics near EPs also pose challenges to the experimental realization~\cite{Ramezanpour2021Genera}. To address these issues, concepts of exceptional lines~\cite{Xu2017} and even exceptional surfaces~\cite{Wiersig2023} have been proposed, offering larger areas in parameter space that may facilitate experimental implementations~\cite{Liu2022,Qiu2022}. In a recent study, Ramezanpour and Bogdanov demonstrated that Kerr nonlinearity can alter the location of EPs in parameter space and proposed the use of Kerr-type nonlinearity to compensate for experimental imperfections, such as fabrication-induced energy level mismatches~\cite{Ramezanpour2021tun}.  This investigation highlights the potential of leveraging intrinsic material nonlinearity as a tool for system control.

Motivated by Ramezanpour and Bogdanov's investigation~\cite{Ramezanpour2021tun}, we in this work study the possibility of tuning the location of EPs in parameter space using a saturable nonlinearity. In contrast to Kerr nonlinearity, saturable nonlinearity possesses unique characteristics as it reaches a maximum saturation value instead of infinitely increasing with field
strength. This type of nonlinearity is prevalent in physical platforms, especially in optical systems~\cite{
Kip2004,Discrete saturable lattice}. The effect of nonlinear saturation has been investigated in various contexts~\cite{Konotop2016,Torner2011}. For example, in a system of two-coupled semiconductor-based resonators, the effect of nonlinear saturation on lasing around an EP was identified as an interesting phenomenon in Ref.~\cite{Hassan2015}: depending on whether the initial lasing occurs in broken or unbroken PT-symmetric modes, saturable nonlinearity can shift the system from the nonlinear PT-broken phase to the unbroken phase or maintain in the unbroken phase. \textcolor{red}{In an exciton-polariton condensate, the saturable gain can modulate the EPs in parameter space~\cite{Wingenbach2023}}.

In Ref.~\cite{Ramezanpour2021tun}, the effect of the Kerr nonlinearity on the existence of EPs was analyzed by calculating the nonlinear energy spectrum, from which the information about the EPs is extracted. A straightforward self-consistent numerical method was employed for these calculations. However, it was noted that this method always does not converge to the target solutions, particularly around the EPs, where the numerical divergence becomes serious~\cite{Ramezanpour2021tun}.  Due to the complexity of the saturable nonlinearity, results from the self-consistent numerical calculation are even less reliable than with the Kerr nonlinearity. We have developed an alternative numerical approach to calculate the nonlinear spectrum. In the calculation, we explicitly incorporate the normalization condition as an auxiliary equation to solve alongside the nonlinear dimer equations using the standard Newton relaxation method. Additionally, we develop an analytical approach to determine the nonlinear spectrum. Both the numerical and analytical approaches yield identical results. From the calculated nonlinear spectrum, we can identify the locations of the EPs and demonstrate that the saturable nonlinearity can adjust the locations of the EPs. Our results not only corroborate those of Ref.~\cite{Ramezanpour2021tun} but also suggest that saturable nonlinearity could serve as a means to compensate for experimental imperfections to facilitate the observation and applications of EPs. 

\section{EPs in a nonlinear dimer}
\label{Sec1}

We consider a physical system comprising two coupled nonlinear resonators~\cite{Ramezanpour2021tun}, described by a nonlinear dimer model with saturable nonlinearity~\cite{Hassan2015}. This system is governed by the following nonlinear equations 
\begin{equation} 	
\left(
\begin{array}{cc}
\delta-2i\Gamma+\mathcal{S}_\text{non}^{(1)} & J \\
J & -\delta+\mathcal{S}_\text{non}^{(2)}
\end{array}
\right)
\left(
\begin{array}{l}
\psi_1 \\
\psi_2
\end{array}
\right)
=\mu
\left(
\begin{array}{l}
\psi_1 \\
\psi_2
\end{array}
\right).
\label{eq:Hamiltonian}
\end{equation}
where $\psi_{1,2}$ represent the mode amplitudes, $\mu$ is the chemical potential, $\delta$ denotes the detuning between two modes, and $J$ is the coupling. The system becomes non-Hermitian due to the finite loss $\Gamma>0$ in the first mode. \textcolor{red}{The loss in the second mode is negligible}. The saturable nonlinearity is expressed as~\cite{Hassan2015}
\begin{equation}
\mathcal{S}_\text{non}^{(j)}=\frac{g_j}{1+|\psi_j|^2},
\end{equation}
with $j=1,2$ and $g_j$ being the nonlinear coefficients. Note that we assume saturable interactions within each mode and neglect the cross-modal interactions~\cite{Hassan2015}.

\subsection{EPs in the linear spectrum}

\begin{figure}[t!]
\centerline{
\includegraphics[width=0.4\textwidth]{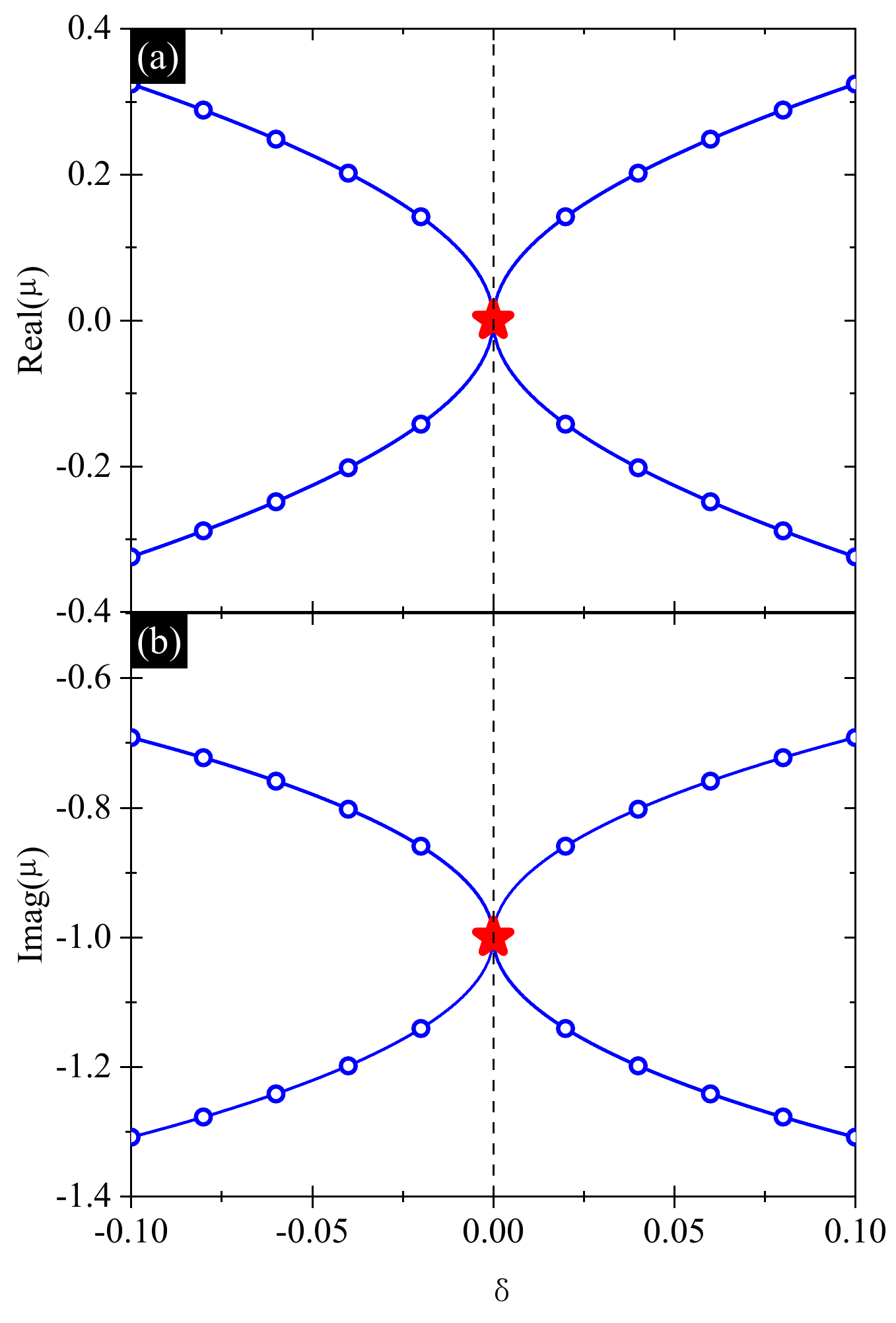}}
\caption{The EP in the linear spectrum ($ g_1=g_2=0 $).
Real and imaginary parts of the eigenvalues 
are shown in (a) and (b), respectively,  with the parameters $J=  \Gamma=1 $. The solid lines are obtained from Eq.(\ref{eq:linearspectrum}) and circles are from solving  Eq.(\ref{coupled}) and Eq.(\ref{renormalization}) using the Newton relaxation numerical method. The red asterisks indicate the location of the EP in Eq.(\ref{linearlocation}), which appears at $\delta=0$ (black dashed vertical line).}
\label{fig:fig1}
\end{figure}

In the absence of the nonlinearity, i.e., $g_1=g_2=0$,  the system becomes a linear passive non-Hermitian system. Such systems have been extensively studied in optics, atomic physics, and acoustics to demonstrate the occurrence of EPs. The two eigenvalues of the linear system can be directly obtained as 
\begin{equation}
    \mu_{\pm}= -i\Gamma \pm \sqrt{J^2+(\delta-i\Gamma)^2}.
    \label{eq:linearspectrum}
\end{equation}
EPs arise when the two eigenvalues become degenerate, i.e., $\mu_+=\mu_-$, which requires $\sqrt{ J^2+(\delta-i\Gamma)^2  }=0$. Therefore, the EPs are located at
\begin{equation}
   \delta_\text{EP}=0, \ \ \  J_\text{EP}=\pm\Gamma, \ \ \ \mu_\text{EP}=-i\Gamma,
   \label{linearlocation}
\end{equation}
in parameter space. This indicates that the precise realization of the EPs demands exact zero detuning and matching between $J$ and $\Gamma$. Figure~\ref{fig:fig1} depicts the two eigenvalues as a function of detuning with parameters set to $J=\Gamma=1$. It shows that both the real and imaginary parts of the chemical potential are degenerate at $\delta=0$, indicating the appearance of the EP. The location of the EP, as defined in Eq.~(\ref{linearlocation}), is labeled by red asterisks. \textcolor{red}{For a very small detuning}, $\delta\ne 0$, there will be gap openings in both the real and imaginary parts of the chemical potential near the EP. The energy splitting is given by $\Delta\mu =|\mu_+ - \mu_-| \approx 2\sqrt{2\Gamma}\sqrt{|\delta|}$. Compared to the Hermitian system\textcolor{red}{~\cite{DP and EP sensitivity}}, the square-root detuning dependence results in a more pronounced energy splitting that could be beneficial for enhanced sensing applications.

\subsection{EPs in the nonlinear spectrum: an analytical approach}

In the presence of the saturable nonlinearity, one cannot directly solve Eq.(\ref{eq:Hamiltonian}) directly to find the nonlinear spectrum. The nonlinearity can introduce many new solutions absent in the linear counterpart, which makes it difficult to identify EPs in the nonlinear spectrum. 

There are three unknown quantities, $\psi_1$, $\psi_2$ and $\mu$, in the two coupled Eqs.(\ref{eq:Hamiltonian}).  To solve them, a third auxiliary equation for these quantities is necessary. The auxiliary equation that we adopt is the normalization condition,
\begin{equation}
    |\psi_1|^2+|\psi_2|^2 = 1.
    \label{normalization}
\end{equation}
which signifies that the particle number, $|\psi_1|^2+|\psi_2|^2$, remains constant for all nonlinear spectra. The normalization condition inspires us to define another useful quantity, the population imbalance $s$, as
\begin{equation}
    s= |\psi_1|^2-|\psi_2|^2.
    \end{equation}
With the normalization condition and the population imbalance, we find $|\psi_1|^2=(1+s)/2$ and $|\psi_2|^2=(1-s)/2$. Substituting them into Eq.(\ref{eq:Hamiltonian}) and cancelling the linear terms related to $\psi_1$ and $\psi_2$, we immediately obtain,
\begin{equation}
    \mu= \frac{\delta}{s}+\frac{g_1(1+s)}{s(3+s)}-\frac{g_2(1-s)}{s(3-s)} -i\Gamma (1+s).
    \label{chemicalpotential}
\end{equation}
Therefore, the nonlinear spectrum $\mu$ will be known once $s$ is obtained. The functional relation between $\mu(s)$ and $s$ is the particular reason why we introduce population imbalance as an extra quantity. Substituting Eq.(\ref{chemicalpotential}) into Eq.(\ref{eq:Hamiltonian})
and solving the determinant of the resulting equations, we get an eighth-order polynomial equation for the population imbalance,
\begin{equation} 	
a_8 s^8+a_6 s^6+a_5 s^5+a_4 s^4+a_3 s^3+a_2 s^2+a_1 s+ a_0=0. \label{eq:8th}
\end{equation}
where
\begin{eqnarray}
    a_8 &=& -\Gamma^2, \notag \\
    a_6 &=& -\delta^2+19\Gamma^2-J^2, \notag \\
    a_5 &=& -2(g_1+g_2)\delta, \notag \\
    a_4 &=& 19\delta^2+6\delta(g_1-g_2) -99\Gamma^2+18J^2-(g_1+g_2)^2, \notag \\
    a_3 &=& 2(g_1+g_2)[10\delta + 3(g_1-g_2)], \notag \\
    a_2 &=& -99\delta^2+20g_1g_2+81\Gamma^2-81J^2-60\delta g_1+60\delta g_2 \notag \\
        & & -8g_1^2-8g_2^2, \notag \\
    a_1 &=& -6(g_2+g_1)(3\delta+g_1-g_2),
    \notag \\
    a_0 &=& 9(3\delta+g_1-g_2)^2. \notag
\end{eqnarray}
By finding the roots of the polynomial equation, we obtain the population imbalance $s$, and subsequently the nonlinear spectrum $ \mu(s)$. Since the population imbalance must be real-valued, only the real roots are physically acceptable. We have checked different parameter regimes and found that the number of real roots may be two, three, or four.

\begin{figure}[t!]
\centerline{
\includegraphics[width=0.4\textwidth]{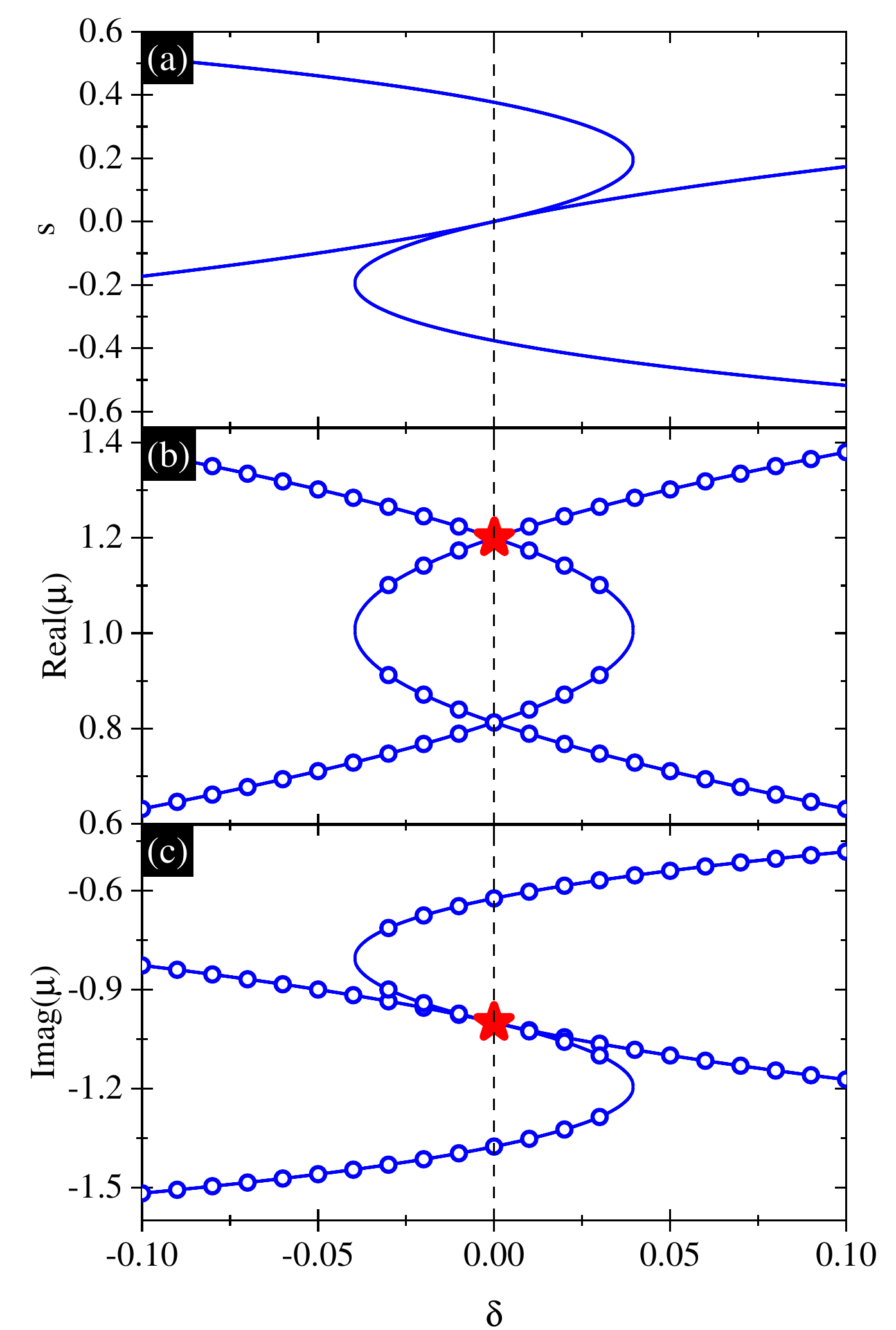}}
\caption{ The EP in the nonlinear spectrum for an equal saturable nonlinearity, 
$ g_1=g_2=1.8 $. The other parameters are $ \Gamma=J=1 $.
(a) The population imbalance $s$ as a function of the detuning, (b) and (c) show the real part and imaginary part of the nonlinear spectrum as a function of the detuning, respectively. The solid lines are from the calculation of Eq.(\ref{chemicalpotential}) and Eq.(\ref{eq:8th}), and circles represent results obtained by solving Eq.(\ref{coupled}) and Eq.(\ref{renormalization}) using the Newton relaxation method. The red asterisks label the location of the EP in Eq.(\ref{nonlinearlocation}).}
\label{fig:fig2}
\end{figure}

Figure~\ref{fig:fig2} demonstrates typical results of the population imbalance $s$ and the spectrum $\mu(s)$ for an equal saturable nonlinearity, $g_1=g_2=1.8$.  Without the nonlinearity, there are two eigenvectors for a given detuning, which results in two different population imbalances. The nonlinearity generates more nonlinear eigenvectors that are linearly independent. Therefore, the population imbalance $s$ may take more than two possible values. In Fig.~\ref{fig:fig2}(a), it is shown that for $|\delta| < 0.038$ there are four values for the population imbalance. Two of them are equal to $s=0$ at $\delta=0$, associated with a symmetry $(\delta, s)\rightarrow (-\delta, -s)$ of Fig.~\ref{fig:fig2}(a). Such symmetry is related to the fact that the coefficients of the odd-order terms are linearly proportional to the detuning when $g_1=g_2$, i.e., $a_{1,3,5}\propto \delta$, and the coefficients of even-order terms are a function of $\delta^2$. 

The extra nonlinear eigenvectors lead to a dramatic change in the nonlinear spectrum $\mu$. As shown in Fig.~\ref{fig:fig2}(b), a loop structure appears in the real part of the nonlinear spectrum when the detuning $|\delta| < 0.038$. Similar to the linear spectrum in Fig.~\ref{fig:fig1}(a), the real part is still symmetric with respect to $\delta=0$. This is related to the symmetry $(\delta, s)\rightarrow (-\delta, -s)$ that dictates $\text{Real}(\mu) [\delta,s]= \text{Real}(\mu) [-\delta,-s]$ when $g_1=g_2$ in Eq.(\ref{chemicalpotential}). In contrast, the imaginary part of the nonlinear spectrum loses such symmetry [see Fig.~\ref{fig:fig2}(c)], since it is linearly proportional to $s$ as shown in Eq.(\ref{chemicalpotential}). There are two degenerate points in the real part of the nonlinear spectrum in Fig.~\ref{fig:fig2}(b). However, only the upper one (labeled by a red asterisk) corresponds to the EP, since the corresponding imaginary part of the chemical potential is also degenerate (labeled by a red asterisk in Fig.~\ref{fig:fig2}(c)). It is interesting to note that the EP is located at $\delta=0$ which is the same as the linear case and $s=0$ at the EP. The equal saturable nonlinearity can not adjust the location of the EP in the detuning space but displaces the nonlinear spectrum.

\begin{figure}[t!]
\centerline{
\includegraphics[width=0.4\textwidth]{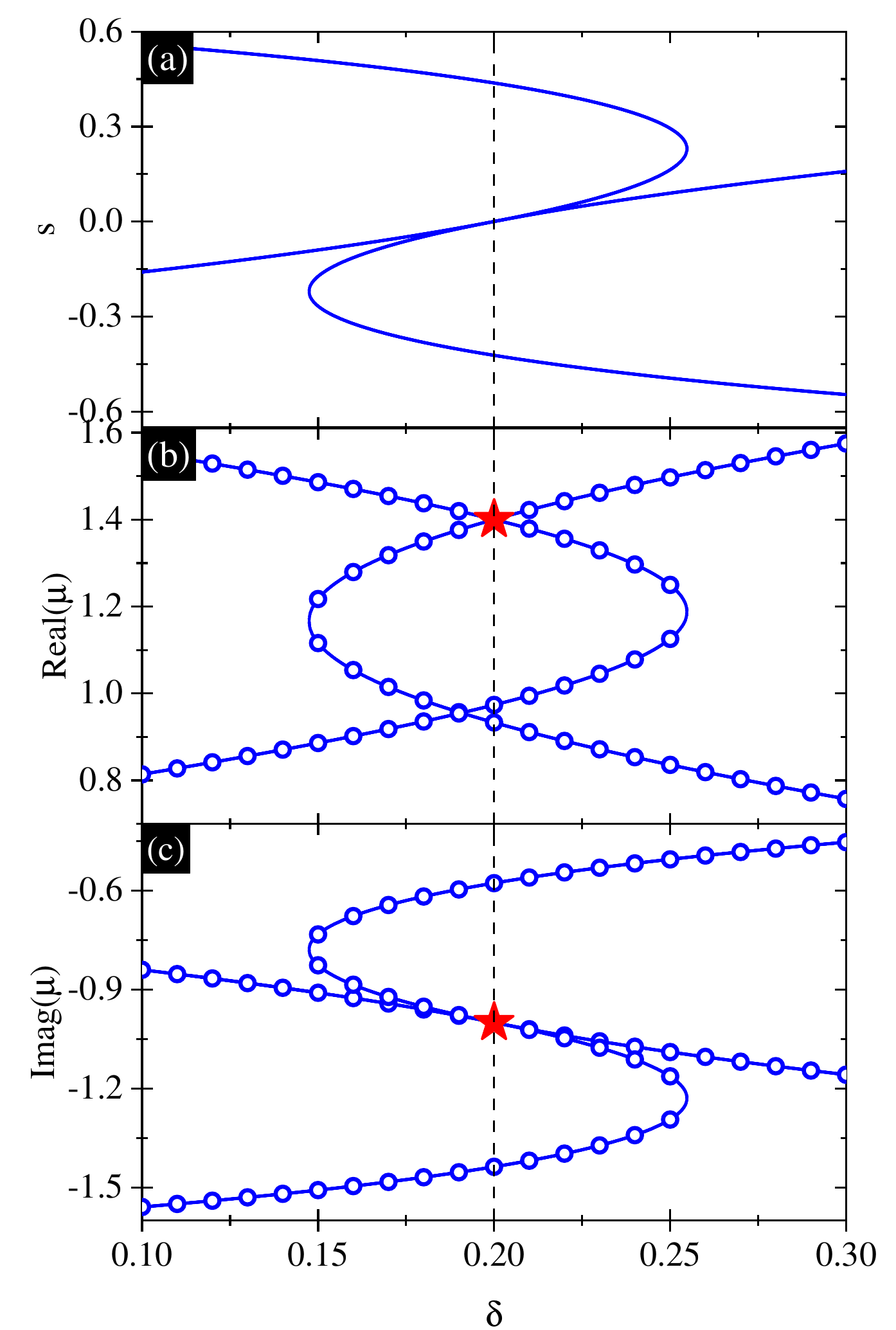}}
\caption{ The EP in the nonlinear spectrum for a non-equal saturable nonlinearity, $ g_1=1.8$ and $g_2=2.4 $. The other parameters are $ \Gamma=J=1 $.
(a) The population imbalance $s$ as a function of the detuning, (b) and (c) show the real part and imaginary part of the nonlinear spectrum as a function of the detuning, respectively. The solid lines are from the calculation of Eq.(\ref{chemicalpotential}) and Eq.(\ref{eq:8th}), and circles represent results obtained by solving Eq.(\ref{coupled}) and Eq.(\ref{renormalization})  using the Newton relaxation method. The red asterisks label the location of the EP in Eq.(\ref{nonlinearlocation}).}
\label{fig:fig3}
\end{figure}

Figure~\ref{fig:fig3} demonstrates typical results of the population imbalance $s$ and the spectrum $\mu(s)$ for a non-equal saturable nonlinearity, $g_1=1.8$ and $g_2=2.4$. The structure of $s$ is similar to that for the equal saturable nonlinearity [see Fig.~\ref{fig:fig3}(a)]. But the structure of $s$ losses the symmetry $(\delta, s)\rightarrow (-\delta, -s)$, since the coefficients of some terms in Eq.(\ref{eq:8th}),  e.g., $a_4$, $ a_3$, $a_2$, $a_1$, and $a_0$, are displaced by $g_1-g_2$.  A nonlinear loop still appears in the real part of the nonlinear spectrum as shown in Fig.~\ref{fig:fig3}(b). There are two degenerate points, but only the upper one (labeled by a red asterisk) is the EP since the corresponding imaginary part of the chemical potential is also degenerate as shown in Fig.~\ref{fig:fig3}(c). Different from the equal saturable nonlinearity, the EP is located at a nonzero detuning, whose value depends on  $g_1-g_2$.

We emphasize that only a non-equal saturable nonlinearity can adjust the location of the EP. This is similar to the case of a Kerr nonlinearity as discovered in Ref.~\cite{Ramezanpour2021tun}. In the following, we provide an analytical approach to find the location of the EP and gain physical insights for the understanding of the results in Figs.~\ref{fig:fig2} and~\ref{fig:fig3}.

 The non-Hermitian Hamiltonian, corresponding to Eq.~(\ref{eq:Hamiltonian}), can be rewritten as
\begin{align}
H &=
\left(
\begin{array}{cc}
\delta-2i\Gamma+\mathcal{S}_\text{non}^{(1)} & J \\
J & -\delta+\mathcal{S}_\text{non}^{(2)}
\end{array}
\right) 
= 
A_0 \mathbf{I}+ H^\text{eff},
\end{align}
where, \textcolor{red}{ $\mathbf{I}$ is the $2\times 2$ identity matrix,} and
\begin{equation}
H^\text{eff}= X\sigma_z+J\sigma_x,
\label{spindependent}
\end{equation}
with 
\begin{eqnarray}
 A_0 &= & -i\Gamma+\frac{1}{2}\frac{g_1}{1+|\psi_1|^2}+\frac{1}{2}\frac{g_2}{1+|\psi_2|^2}, \notag  \\
 X & =	& \delta-i\Gamma+\frac{1}{2}\frac{g_1}{1+|\psi_1|^2}-\frac{1}{2}\frac{g_2}{1+|\psi_2|^2}, \notag 
\end{eqnarray}
and $\sigma_{x,z}$ are the Pauli matrices. The eigenvalues of $H$ can be formally expressed as
\begin{eqnarray}
\mu_\pm = A_0 \pm \sqrt{X^2+J^2}. 
\label{nonlinearspectrum}
\end{eqnarray}
EPs, if exist, should appear at the degeneracy, $\mu_+=\mu_-$, which requires $X^2+J^2=0$. This requirement leads to,
\begin{equation}
    J= \pm \Gamma,
\end{equation}
 \begin{equation}
\delta+\frac{1}{2}\frac{g_1}{1+|\psi_1|^2}-\frac{1}{2}\frac{g_2}{1+|\psi_2|^2} = 0. \label{eq:nonlinear}
 \end{equation}
Substituting these results into Eq.(\ref{spindependent}), we find $H^\text{eff}$ at the EPs as, 
\begin{equation}
H_\text{EP}^\text{eff}=  J(\mp i \sigma_z+\sigma_x),
\end{equation}
where, $\mp$ corresponds to $\Gamma=\pm J$. One can immediately find that the eigenvalues of
$H_\text{EP}^\text{eff} $ are degenerate and are zero. Meanwhile, there is only one eigenvector as
\begin{equation}
   \begin{pmatrix}
       \psi_1\\ \psi_2
   \end{pmatrix}_\text{EP} = \frac{1}{\sqrt{2}} \begin{pmatrix}
       1\\ \pm i 
   \end{pmatrix}.
\end{equation}
This is because at the EPs the two eigenvectors coalesce into one.  The eigenvector of the EP satisfies the normalization condition $|\psi_1|^2+|\psi_2|^2=1$. It is interesting to note that the population imbalance at the EPs is always zero, i.e., $s_\text{EP}=|\psi_{1\text{EP}} |^2-|\psi_{2\text{EP}} |^2=0$. This explains why the population imbalance vanishes at the EPs in Fig.~\ref{fig:fig2} and Fig.~\ref{fig:fig3}.  Substituting the eigenvector of the EPs into Eq.(\ref{nonlinearspectrum}) and Eq.(\ref{eq:nonlinear}), we find the location of the EP,
\begin{eqnarray}
\delta_\text{EP} &=& -\frac{1}{3}(g_1-g_2), \notag \\	
J_\text{EP} &=& \pm \Gamma,  \label{nonlinearlocation} \\
\mu_\text{EP} &= &-i\Gamma+\frac{1}{3}(g_1+g_2) \notag .  
\end{eqnarray}
These analytical results are labeled by red asterisks in Figs.~\ref{fig:fig2} and \ref{fig:fig3}, where they excellently agree with the results obtained by solving the polynomial equation in Eq.(\ref{eq:8th}). The adjustment of EPs by saturable nonlinearities manifests in two primary aspects: the non-equal saturable nonliterary can tune the location of the EP in the detuning space; only the real part of the nonlinear spectrum at the EP is affected by the nonlinearity.  At last, we emphasize that whether there is nonlinearity or not the appearance of the EPs requires $J=\pm \Gamma$. If $J\ne \pm \Gamma$, EP will not appear. This is why in Figs.~\ref{fig:fig2} and \ref{fig:fig3} we use $J=\Gamma=1$ as typical examples. 

\subsection{EPs in the nonlinear spectrum: the Newton relaxation numerical method}
\label{Sec2}
Over the last few years, several methods have been developed to numerically find the eigenvalues of the non-Hermitian Hamiltonian in the presence of nonlinearities. These include the polar representation~\cite{Li2011,Suwunnarat2020}, stokes parameters and Bloch sphere~\cite{Graefe2012,Graefe2010,Ramezani2010}, angular momentum operators~\cite{Graefe2008} and analytic continuation~\cite{Dast2013_1,Dast2013_2}. In recent studies~\cite{Ramezanpour2021tun,Wang2022,Wu2022,Bai2023}, a self-consistent method was utilized to numerically find the eigenvalues of a non-Hermitian Hamiltonian in the presence of Kerr nonlinearities.  However, the self-consistent numerical method may have an unsatisfactory convergence around the EPs~\cite{Ramezanpour2021tun}. Here, we develop an alternative approach to numerically solve Eq.(\ref{eq:Hamiltonian}) using the standard Newton relaxation method. This fully numerical approach offers a means to validate the accuracy of the polynomial equation method presented in the previous section.

In our approach, Eq.(\ref{eq:Hamiltonian}) together with the auxiliary equation of the normalization condition in Eq.(\ref{normalization}) should be numerically solved.  We note that there is a gauge freedom in Eq.(\ref{eq:Hamiltonian}) and Eq.(\ref{normalization}): if $\psi=(\psi_1,\psi_2)^T$ is an eigenvector of Eq.(\ref{eq:Hamiltonian}) then the states $e^{i\phi}\psi$ with $\phi$ being an arbitrary real constant are also eigenvectors. The global phase $\phi$ is trivial as it does not affect any physical results. This stimulates us to choose a fixed gauge $\phi$. With such a pre-selected gauge, the mode amplitude $\psi_1$ becomes purely real-valued.  Thus, it is eligible to assume the following ansatz for the order parameters
\begin{eqnarray}
    \psi_1 &=& \phi_1,  \notag  \\
    \psi_2 &=& \phi_2 + i\phi_3,
\end{eqnarray}
where $\phi_{1,2,3}$ are real-valued. Consequently, the normalization condition becomes
\begin{equation}
    \phi_1^2+\phi_2^2+\phi_3^2 = 1.
    \label{renormalization}
\end{equation}
We further assume  $\mu=\mu_r+i\mu_i$ with $\mu_r$ and $\mu_i$ being the real and imaginary parts of the chemical potential, respectively. Substituting the ansatz for the order parameter and $\mu$ into Eq.~(\ref{eq:Hamiltonian}), we obtain the following coupled nonlinear equations, 
\begin{eqnarray}
(\delta -\mu_r+\frac{g_1}{1+\phi_1^2})\phi_1 +J \phi_2 & =0, \notag 
\\
-(2\Gamma +\mu_i )\phi_1 +J \phi_3  & =0,  \notag \\
J\phi_1+(-\delta -\mu_r +\frac{g_2}{1+\phi_2^2+\phi_3^2})\phi_2 +\mu_i \phi_3& =0, \notag \\
(\delta +\mu_r-\frac{g_2}{1+\phi_2^2+\phi_3^2})\phi_3+\mu_i \phi_2&=0.
\label{coupled}
\end{eqnarray}
In the above coupled nonlinear equations, all quantities are real. The five real parameters $(\phi_1, \phi_2, \phi_3, \mu_r, \mu_i)$ can be determined from the coupled nonlinear equations in Eq.(\ref{coupled}) and Eq.(\ref{renormalization}) using the standard Newton relaxation method.  We first apply this numerical approach to calculate the linear spectrum. The results are described by circles in Fig.~\ref{fig:fig1}. It shows that the numerical results exactly agree with the analytical results in Eq.(\ref{eq:linearspectrum}). Then we apply it to find the nonlinear spectrum in Figs.~\ref{fig:fig2} and~\ref{fig:fig3} (with the results represented by circles in the figures). The perfect agreement between the results from the numerical method and the analytical approach based on the polynomial equation gives a solid validation of the discussions in the previous section.

\section{Conclusions}
In summary, we have discussed the possibility of adjusting the location of the EPs in parameter space by a saturable nonlinearity. The system we consider consists of two coupled resonators with saturable nonlinearity.  A finite loss in one resonator brings the system non-Hermitian. Consequently, the non-Hermitian system may possess EPs. We have developed a polynomial equation for the population imbalance to calculate the nonlinear spectrum, from which the EPs are identified. Furthermore, we derive the effective Hamiltonian at the EPs which allows us to identify the location of the EPs analytically. We find that only the non-equal saturable nonlinearity can adjust the location of the EP in the detuning space. Finally, we develop a fully numerical approach for the calculation of the nonlinear spectrum. The application of the numerical method validates the polynomial equation method.  

Our findings showcase the significant impact of saturable nonlinearity on the eigenvalues of a non-Hermitian dimer system. The nonlinearity not only introduces more than two eigenvalues but also alters the location of the EPs, offering a valuable avenue for tuning the EPs in sensing applications and experimental implementations.

\begin{acknowledgments}
This work is supported by National Natural Science Foundation of China with Grants No. 12374247 and No. 11974235. C.Q. is supported (in part) by ACC-New Jersey under Contract No. W15QKN-18-D-0040.

\end{acknowledgments}


\end{document}